\documentclass[prd,aps,reprint,superscriptaddress, nofootinbib,amsmath,amssymb]{revtex4-2}
\usepackage[dvipsnames]{xcolor}
\usepackage{dcolumn}
\usepackage{comment}
\usepackage{graphicx}
\usepackage{hyperref}
\usepackage{mathrsfs}
\usepackage{tabularx}
\usepackage{multirow}
\usepackage[normalem]{ulem}
\hypersetup{colorlinks=true,linkcolor=blue}
\newcolumntype{Y}{>{\centering\arraybackslash}X}

\begin{document}
\title{Dissipative Unimodular Gravity: Linking Energy Diffusion to Bulk Viscosity as an Alternative to $\Lambda$CDM under DESI DR2 Data}

\author{Norman Cruz}
\email{norman.cruz@usach.cl}
\affiliation{Departamento de Física, Universidad de Santiago de Chile, Avenida Victor Jara 3493, Estación Central, 9170124 Santiago, Chile}
\affiliation{Center for Interdisciplinary Research in Astrophysics and Space Exploration (CIRAS), Universidad de Santiago de Chile, Avenida Libertador Bernardo O’Higgins 3363, Estación Central, 9170022 Santiago, Chile}

\author{Esteban González}
\email{esteban.gonzalez@ucn.cl}
\affiliation{Departamento de Física, Universidad Católica del Norte, Avenida Angamos 0610, Casilla 1280, 1270709 Antofagasta, Chile}

\author{Guillermo Palma}
\email{guillermo.palma@usach.cl}
\affiliation{Departamento de Física, Universidad de Santiago de Chile, Avenida Victor Jara 3493, Estación Central, 9170124 Santiago, Chile}

\begin{abstract}
In this paper, we perform a theoretical and observational study of the presence of viscosity in the Unimodular Gravity formalism, a pioneering approach that, to the best of our knowledge, has not been previously considered within the present context. Specifically, we study a flat FLRW universe at late times, where matter experiences dissipative processes in the form of a bulk viscosity, in the framework of Eckart's theory, which is linked to the energy diffusion function $Q$ through the power law $\xi=\xi_{0}\left|Q\right|^{1/2}$, being $\xi_{0}$ a positive dimensionless parameter. By assuming the Ansatz $Q=\nu H^{2}$, where $H$ is the Hubble parameter and $\nu$ is a dimensionless arbitrary constant, we find analytical solutions for the cosmological evolution. We test these models against the most recent cosmological observations, including type Ia supernovae, baryon acoustic oscillations, cosmic chronometers, gravitational lensing, and black hole shadow data. Our results show that two of the tested models provide a significantly better fit to the data ($\chi_{\text{min}}^{2}$) and remain as competitive as the $\Lambda$CDM model according to the Bayesian Information Criterion. These findings, combined with the inherent ability of Unimodular Gravity to alleviate the cosmological constant problem, position dissipative UG as a robust and compelling alternative to the standard model, potentially suggesting that a very small but nontrivial energy nonconservation is compatible with the late-time observational data.
\end{abstract}
\maketitle

%%%%%%%%%%%%%%%%%%%%%%%%%%%%%%%%%%%%%%%%%%%%%%
\section{\label{sec:Introduction}Introduction}
%%%%%%%%%%%%%%%%%%%%%%%%%%%%%%%%%%%%%%%%%%%%%%
The standard $\Lambda$CDM ($\Lambda$ Cold Dark Matter) model, a spatially flat universe described by the Friedman-Lemaître-Robertson-Walker (FLRW) metric, with a total matter component consisting of approximately $30\%$ dark matter (DM) and $70\%$ dark energy (DE)~\cite{Planck:2018vyg}, has successfully described many observational results, such as those obtained from type Ia supernovae (SNe Ia)~\cite{SupernovaSearchTeam:1998fmf,SupernovaCosmologyProject:1998vns}, the cosmic microwave background (CMB)~\cite{SDSS:2003eyi}, information from large-scale structure (LSS) formation~\cite{2013ApJS..208...19H}, among others. However, this model is not without problems, with one of them being the most significant theoretical challenge of all time, known as the cosmological constant (CC) problem. This issue involves an enormous disparity of 120 orders of magnitude between the value of the CC derived from observational data and the value predicted theoretically by quantum field theory~\cite{Weinberg:1988cp,Padmanabhan:2002ji,Nobbenhuis:2004wn,Burgess:2013ara,Padilla:2015aaa}.

One of the approaches to tackle the aforementioned problem is the unimodular gravity (UG) theory, whose concept dates back to the era of general relativity (GR), with the first work on this topic published by Einstein himself~\cite{Einstein:1919gv,lorentz1952principle}. In principle, Einstein wrote the traceless equations of GR for a metric with a fixed determinant ($|g|=1$) as a convenient way to partially fix a coordinate system to simplify some calculations in certain situations~\cite{Bufalo:2015wda}. Nowadays, the definition of UG is usually based on invariance under a restricted group of diffeomorphisms, in which the determinant of the metric can be set equal to a fixed scalar density $\epsilon_{0}$, of the form $|g|=\epsilon_{0}$~\cite{Bufalo:2015wda}, which provides a fixed-volume element in spacetime. A consequence of this invariance is that, classically, UG produces the same physics as GR, with the field equation for the metric being the traceless Einstein equation or, owing to the Bianchi identity, the Einstein equation with a CC~\cite{Unruh1988in}. However, the key distinction between GR and UG is that, in the latter, the CC appears as an integration constant rather than a coupling constant. The benefit is that the vacuum energy density of quantum fields can be removed from the field equations by rescaling an additional component of the energy-momentum tensor that arises from the restricted invariance, which leads to UG not suffering from the CC problem~\cite{Smolin:2009ti}. Another important property of UG, caused by the restricted diffeomorphism invariance, is the appearance of energy diffusion, characterized by a specific energy diffusion function $Q$. This function opens new windows for exploring physics beyond the $\Lambda$CDM model. For example, in Ref.~\cite{Corral:2020lxt}, a cosmological UG model has been explored using a $Q$ function described by the barotropic model $Q \equiv x\rho$, where $x \in [-1,1]$ is a constant similar to the barotropic index $\omega$ of a barotropic fluid, and $\rho$ represents the energy density of the main fluid. This model successfully describes the joint analysis of SNe Ia and observational Hubble parameter data (OHD) in comparison with the $\Lambda$CDM model, providing insights that a very small but nontrivial energy nonconservation is compatible with the model. On the other hand, in Ref.~\cite{Garcia-Aspeitia:2019yni}, a diffusion function $Q$ that considers third-order derivatives of the scale factor, related to a jerk parameter, is used to describe the late accelerated expansion without the inclusion of a DE component. Also, interaction between DM and DE within the framework of UG to address other tensions in the $\Lambda$CDM model, such as the $H_0$ tension, has been investigated in~\cite{Landau:2022mhm,LinaresCedeno:2020uxx}.

Other approaches are viscous cosmologies, which appear as natural extensions to tackle the problems associated with the $\Lambda$CDM model. The most important in recent years is the $H_{0}$ tension mentioned above, which presents a discrepancy of $5\sigma$ between the measurements obtained from Planck CMB (inferred from the $\Lambda$CDM model) and those obtained from local measurements of Cepheids (model-independent)~\cite{Riess:2021jrx}. This tension is also supported by the gravitational lensing measurements of the H0LiCOW collaboration, with a discrepancy of $5.3\sigma$ with respect to the value inferred from Planck CMB~\cite{Wong:2019kwg}. Another tension is the one obtained from the global EoR Signature (EDGES) experiment, which detects an excess of radiation at $z \approx 17$ in the reionization epoch, not predicted by the $\Lambda$CDM model~\cite{Bowman:2018yin}. Both challenges are addressed by extending the nature of DM to include bulk viscosity, a topic that has been extensively reviewed in~\cite{Brevik:2017msy} regarding its various cosmological applications~\cite{Bamba:2012cp}. Specifically, the $H_{0}$ tension is addressed in~\cite{Elizalde:2020mfs,Normann:2021bjy}, whereas the EDGES results are explained in~\cite{Bhatt:2019qbq}. As in the UG case with the diffusion function, dissipative cosmologies open new windows for exploring physics beyond the $\Lambda$CDM model. In this regard, it has been recently proposed that the bulk viscosity can be interpreted as a natural consequence of the velocity gradient of the comoving expansion, providing a novel mechanism for its origin while ensuring that the first and second laws of thermodynamics are satisfied at the apparent horizon~\cite{Paul:2025rqe}. In the case of unified models, where bulk viscous DM causes the accelerated expansion of the universe due to the negativeness of the viscous pressure, DE can be avoided, in principle, to overcome the CC problem. Nevertheless, for models where the bulk viscosity is only a function of the DM density, it has been shown in~\cite{Palma:2024qrw} that, regardless of the chosen parametrization for the bulk viscosity, very high values of this parameter are required to achieve an accelerated phase at late times, which are not expected due to the present very low value of DM. Following this line, the authors found that the dynamical system analysis for this unified DM model shows that there is a lack of a complete and consistent description of the background cosmic evolution since the very low viscosity during structure formation, compatible with the data, demands the counterintuitive behavior of bulk viscosity diminishing as the energy density increases. Of course, all of these inconveniences are not present if another form of DE fills the universe. In this sense, in UG, where the DE component appears as an integration constant, the above problems can be avoided.

In Refs.~\cite{Josset:2016vrq, pellecchia2026darkenergygenesismodeling}, the authors state that, at the classical level, UG is the simplest modification of GR that allows for the inclusion of dissipative phenomena. Nevertheless, in the context of an extension of the dark sector that considers a non-perfect DM fluid from the beginning, it is necessary to additionally take into account a theory of dissipative relativistic fluids. Therefore, the aim of this paper is to expand the scope of the UG formalism to encompass viscous processes, a pioneering approach that, to the best of our knowledge, has not been previously explored in this particular context. Following this aim, we propose an approach that relates the energy diffusion function characterizing UG to the viscosity experienced by the matter component of the universe during its cosmic evolution, assuming that this process of non-energy conservation leads to such dissipative effects.

This paper is organized as follows: Section \ref{sec:DissipativeUG} provides a brief description of the UG theory, presenting the general expression for the energy-momentum tensor under dissipative processes within the framework of Eckart's theory, while accounting for a general energy diffusion function Q. Section \ref{sec:Model} presents the equations of motion for a FLRW spacetime at late times, where an analytical solution is derived for a bulk viscosity $\xi$ related to the diffusion function $Q$. Section \ref{sec:Constraints} describes the fitting procedure using observational cosmological data—including type Ia Supernovae, cosmic chronometers, baryon acoustic oscillations, gravitational lensing, and black hole shadows—detailed in Subsections \ref{subsec:SNeIa}--\ref{subsec:BHS}. Subsection \ref{subsec:priors} discusses the free parameters and the priors considered in our MCMC analysis. In Section \ref{sec:Results}, we analyze the obtained results and provide an initial discussion of the model's performance. In Section~\ref{sec:DifussionAndBulk}, we discuss some physical arguments supporting the hypothesis that cosmological dissipation naturally arises from energy non-conservation via diffusion. Finally, Section \ref{sec:Conclusions} summarizes our findings and presents the conclusions of our study.

%%%%%%%%%%%%%%%%%%%%%%%%%%%%%%%%%%%%%%%%%%%%%%%%%%%%%%%%
\section{\label{sec:DissipativeUG}Dissipative Unimodular Gravity}
%%%%%%%%%%%%%%%%%%%%%%%%%%%%%%%%%%%%%%%%%%%%%%%%%%%%%%%%
To obtain the field equations for dissipative UG, we first need to derive the UG version of Einstein's field equations (EFE), following Refs.~\cite{Einstein:1919gv,lorentz1952principle,Bufalo:2015wda,Unruh1988in,Anderson1971pn,vanderBij:1981ym,Buchmuller1988wx,Henneaux1989zc,Ellis:2013uxa,LinaresCedeno:2020uxx}. In this sense, the EFE of GR, without the inclusion of the CC, are given by
\begin{equation}\label{EFE}
     R_{\mu\nu}-\frac{1}{2}Rg_{\mu\nu}=\kappa T_{\mu\nu},
\end{equation}
where $R_{\mu\nu}$ is the Ricci tensor, $R$ the Ricci scalar, $g_{\mu\nu}$ is the metric tensor of the four-dimensional spacetime, $T_{\mu\nu}$ is the total energy-momentum tensor, and $\kappa=8\pi G/c^{4}$ with $G$ being the gravitational constant and $c$ the speed of light. Due to the Bianchi identity, these equations are subject to the following consistency relation:
\begin{equation}\label{nablaG}
    \nabla^{\nu}\left(R_{\mu\nu}-\frac{1}{2}Rg_{\mu\nu}\right)=0,
\end{equation}
which leads to the conservation equation for the energy-momentum tensor
\begin{equation}\label{conservation}
\nabla^{\nu}T_{\mu\nu}=0.
\end{equation}

The EFE have successfully accounted for all observed gravitational phenomena, and they are the accepted foundation for field theories incorporating gravity. Moreover, since the discovery of the recent acceleration in the expansion of the Universe~\cite{{SupernovaSearchTeam:1998fmf,SupernovaCosmologyProject:1998vns}}, the EFE can successfully describe this acceleration by the inclusion of the CC ($\Lambda$) in Eq. \eqref{EFE} of the form
\begin{equation}\label{EFELambda}
     R_{\mu\nu}-\frac{1}{2}Rg_{\mu\nu}+\Lambda g_{\mu\nu}=\kappa T_{\mu\nu},
\end{equation}
a choice that is also supported by different observations~\cite{Planck:2018vyg,Brout:2022vxf}. However, in UG, we can obtain a different origin for the CC by using the trace-free part of the EFE \eqref{EFE}, resulting in the trace-free EFE given by~\cite{Ellis:2013uxa}
\begin{equation}\label{TFEE}
    R_{\mu\nu}-\frac{1}{4}Rg_{\mu\nu}=\kappa\left(T_{\mu\nu}-\frac{1}{4}Tg_{\mu\nu}\right),
\end{equation}
where $T$ is the trace of the energy-momentum tensor. This traceless structure is consistently maintained in the first-order formulation of UG, where the field equations for the vierbein are naturally trace-free and allow for the inclusion of matter with nontrivial spin density and spacetime torsion~\cite{Bonder:2018mfz}. Taking the divergence of Eq. \eqref{TFEE}, and using the Bianchi identity, it is obtained
\begin{equation}\label{nablaTFEE}
    \frac{1}{4}\nabla_{\nu}\left(R+\kappa T\right)=\kappa \nabla^{\mu}T_{\mu\nu}.
\end{equation}
Note that if we define $4\lambda\equiv R+\kappa T$, then we can integrate Eq. \eqref{nablaTFEE}, yielding
\begin{equation}\label{lambda}
    \lambda=\frac{R+\kappa T}{4}=\Lambda+\int_{l}{J}, 
\end{equation}
with $\Lambda$ emerging as an integration constant and $J=\kappa \nabla^{\mu}T_{\mu\nu}$ representing the energy–momentum current violation, which is integrated on some arbitrary path $l$~\cite{LinaresCedeno:2020uxx}. The advantage of obtaining the CC as an integration constant is that the vacuum energy density of quantum fields can be removed from the field equations by rescaling an additional component of the energy-momentum tensor that appears from the restricted invariance~\cite{Smolin:2009ti}, which prevents UG from suffering from the CC problem. Finally, from Eqs. \eqref{TFEE} and \eqref{lambda}, we obtain the UG version of the EFE
\begin{equation}\label{UGEFE}
    R_{\mu\nu}-\frac{1}{2}Rg_{\mu\nu}+\left(\Lambda+\int_{l}{J}\right)g_{\mu\nu}=\kappa T_{\mu\nu}.
\end{equation}

Comparing Eqs. \eqref{EFELambda} and \eqref{UGEFE}, we can see that $\lambda$ plays the role of an effective CC (see Eq. \eqref{lambda}), which can be equal to the CC of GR when $J=0$. This case implies the conservation of the energy-momentum tensor according to Eq. \eqref{conservation}. Nevertheless, this is not generally true in the UG formalism, according to equation \eqref{nablaTFEE}. In other words, in the UG framework, the conservation equation \eqref{conservation} is no longer a direct consequence of the geometrical identity equation \eqref{nablaG} and must be introduced as a separate assumption. Otherwise, there could generally be a nonzero divergence of the energy-momentum tensor. Therefore, from now on, we define the diffusion function $Q$ as
\begin{equation}\label{Intdiffusion}
    Q\equiv\int_{l}{J}=\kappa\int_{l}\nabla^{\mu}T_{\mu\nu},
\end{equation}
which measures the nonconservation of the energy-momentum tensor. It is important to mention that, since the integral in Eq. \eqref{Intdiffusion} is over some arbitrary path $l$, the diffusion function is, in principle, arbitrary.

Another alternative is to obtain the UG version of the EFE through invariance under a restricted group of diffeomorphisms. In this case, we can set the determinant of the metric to a fixed scalar density $\epsilon_{0}$ of the form
\begin{equation}\label{fixedmetric}
    \sqrt{-g}=\epsilon_{0},
\end{equation}
which provides a fixed volume element in spacetime. In GR, the symmetry group is the diffeomorphism group (Diff), $\delta g_{\mu\nu}=\nabla_{\mu}\xi_{\nu}+\nabla_{\nu}\xi_{\mu}$, whereas for UG, the condition \eqref{fixedmetric} requires that~\cite{Bufalo:2015wda}
\begin{equation}\label{UGcondition}
    \delta\sqrt{-g}=\frac{1}{2}\sqrt{-g}g^{\mu\nu}\delta g_{\mu\nu}=0,
\end{equation}
which breaks the diffeomorphism invariance down to transverse diffeomorphisms (TDiff), satisfying $\nabla_{\mu}\xi^{\mu}=0$~\cite{Padilla:2014yea}. The consistency of this restricted symmetry is supported by gauge-theoretical studies showing that the algebra of volume-preserving local translations closes off-shell, even when the volume element is non-dynamical~\cite{Corral:2018hxi}. We can introduce this formalism into the Einstein-Hilbert action by adding a Lagrange multiplier $\lambda$ that fixes the volume element on shell as follows~\cite{Corral:2020lxt}:
\begin{align}\label{UGaction}\notag
 S\left[g_{\mu\nu},\lambda,\Psi\right] &= \frac{1}{2\kappa}\int{d^4 x\sqrt{-g}\left[R-2\lambda\left(1-\frac{\varepsilon_0}{\sqrt{-g}}\right)\right]}\\
  &\quad+\int{d^4 x\sqrt{-g}\,\mathscr{L}_m\left[g_{\mu\nu},\Psi\right]},
\end{align}
where $\mathscr{L}_m$ is the matter Lagrangian, and $\Psi$ represents the matter fields.  Note that, if we perform arbitrary variations on Eq. \eqref{UGaction} with respect to $g_{\mu\nu}$, we obtain
\begin{equation}\label{UGEFEaltern}
    R_{\mu\nu}-\frac{1}{2}R g_{\mu\nu}+\lambda g_{\mu\nu}=\kappa T_{\mu\nu}.
\end{equation}
Taking the trace of Eq.~\eqref{UGEFEaltern}, the variable cosmological constant can be solved algebraically to yield $4\lambda\equiv R+\kappa T$. In addition, a variation with respect to $\lambda$ gives the UG condition \eqref{fixedmetric}. Finally, as it is shown in~\cite{Josset:2016vrq}, the Noether theorem associated with the TDiff symmetry implies, as in~\cite{Corral:2020lxt}, the following result:
\begin{align}\label{noether}
 \nabla^\mu\left(T_{\mu\nu} - g_{\mu\nu}Q \right) = 0,
\end{align}
where $Q$ is the diffusion function given by Eq. \eqref{Intdiffusion}. Note that this equation makes clear that $\nabla^\mu T_{\mu\nu}$ is no longer a conserved quantity, as we saw before.

Having introduced the concept of UG, we will now explore the modifications to the EFE for UG when dissipative processes are considered. This approach aims to offer a more realistic depiction of the fluids that constitute the Universe. If we assume that the matter sector can be modeled as a perfect fluid, its energy-momentum tensor can be expressed as
\begin{equation}\label{energy-momentum}
T_{\mu\nu}=p\,g_{\mu\nu}+\left(\rho+p\right)u_{\mu}{u_\nu},
\end{equation} 
where $p$ and $\rho$ are the pressure and energy density of the fluid, respectively, and $u_{\mu}$ corresponds to the four-velocity of the fluid element. In a locally Cartesian comoving inertial frame, the four-velocity components are defined as $u_{0}=1$ and $u_{i}=0$, with a normalization of the form $\eta_{\alpha\beta}u^{\alpha}u^{\beta}=-1$ in any inertial frame. In addition, $p$ and $\rho$ are related by an equation of state (EoS) of the form $p=\omega\rho$, also known as a barotropic EoS, where $\omega$ is the barotropic index. Some values of this index are $\omega=0$ for CDM and $\omega=1/3$ for radiation. Working in the framework of the relativistic thermodynamic theory of Eckart for non-perfect fluids, we need to introduce a small correction $\Delta T_{\mu\nu}$ to the energy-momentum tensor $T_{\mu\nu}$ in Eq. \eqref{energy-momentum} of the form~\cite{PhysRev.58.919}
\begin{equation}\label{deltaenergy-momentum}
\Delta T_{\mu\nu}=-3H\xi\left(g_{\mu\nu}+u_{\mu}u_{\nu}\right), 
\end{equation}
where $\xi$ is the bulk viscous coefficient, which can particularly depend on the temperature and pressure of the dissipative fluid (further details can be found in~\cite{Weinberg:1972kfs}), and $H$ is the Hubble parameter. It is important to mention that we consider only bulk viscosity due to the homogeneity and isotropy of the Universe at late times, and that the fluid does not experience heat flow. Then, the energy-momentum tensor in a general coordinate system would be 
\begin{equation}\label{Eckartenergy-momentum}
T_{\mu\nu}\rightarrow T_{\mu\nu}+\Delta T_{\mu\nu}=P_{eff}\,g_{\mu\nu}+\left(\rho+P_{eff}\right)u_{\mu}u_{\nu},    
\end{equation} 
where 
\begin{equation}\label{Peff}
    P_{eff}=p+\Pi, 
\end{equation}
is an effective pressure, with $\Pi=-3H\xi$ the viscous pressure of the dissipative fluid. By comparing Eqs. \eqref{energy-momentum} and \eqref{Eckartenergy-momentum}, we can see that the dissipative fluid can be interpreted as a perfect fluid but with an effective pressure that consists of the equilibrium pressure $p$ plus the bulk viscous pressure $\Pi$.

The remarkable feature of this procedure is that, since Eckart's theory only produces a modification on the energy-momentum tensor according to Eq. \eqref{Eckartenergy-momentum}, the UG procedure remains unchanged, because the diffusion function \eqref{Intdiffusion} is an integral over an arbitrary path and, therefore, a modification on the energy-momentum tensor leaves the diffusion function as an arbitrary function. On the other hand, Eckart's theory is constructed by assuming that $\nabla^{\mu}\left(T_{\mu\nu}+\Delta T_{\mu\nu}\right)=0$~\cite{Maartens:1995wt}, which leaves Eq. \eqref{noether} unchanged. Consequently, the results are the standard Friedmann equations for UG with energy diffusion, with the addition of dissipation in the same way as in GR. From now on, units are chosen so that $8\pi G = c = k_{B} = 1$ throughout this manuscript.

%%%%%%%%%%%%%%%%%%%%%%%%%%%%%%%%%%%%%%%%%%%%%%%%%%%%%%%
\section{\label{sec:Model}Late-time Cosmological model}
%%%%%%%%%%%%%%%%%%%%%%%%%%%%%%%%%%%%%%%%%%%%%%%%%%%%%%%
In GR, for a spatially flat FLRW metric with line element
\begin{equation}\label{FLRW}
 ds^2=-dt^{2}+a^{2}(t)\left(dr^{2}+r^{2}d\vartheta^{2}+r^{2}\sin^{2}\left(\vartheta\right) d\varphi^{2}\right),
\end{equation}
where $a(t)$ is the scale factor and $t$ the cosmic time, we can obtain from Eqs. \eqref{EFELambda} and \eqref{energy-momentum} the Friedmann equations that govern the evolution of a universe dominated by a single perfect fluid and a CC, which take the form
\begin{equation}\label{GRfluids}
    3H^{2}=\rho+\Lambda,
\end{equation}
\begin{equation}\label{GRpressures}
    2\dot{H}+3H^2=-p+\Lambda,
\end{equation}
with the conservation equation \eqref{conservation} given by
\begin{equation}\label{GRconservation}
    \dot{\rho}+3H\left(\rho+p\right)=0,
\end{equation}
where $H\equiv\dot{a}/a$ and a ``dot'' denotes the derivative with respect to cosmic time. On the other hand, for a dissipative UG model dominated by a single non-perfect fluid in the framework of Eckart's theory, Eqs. \eqref{GRfluids}, \eqref{GRpressures}, and \eqref{GRconservation} become
\begin{equation}\label{UGfluids}
    3H^{2}=\rho+\Lambda+Q,
\end{equation}
\begin{equation}\label{UGpressures}
2\dot{H}+3H^{2}=-p-\Pi+\Lambda+Q,
\end{equation}
\begin{equation}\label{UGconservation}
    \dot{\rho}+3H\left(\rho+p+\Pi\right)=-\dot{Q},
\end{equation}
which can be obtained from Eqs. \eqref{UGEFE}, \eqref{Intdiffusion}, \eqref{Eckartenergy-momentum}, and \eqref{FLRW}. Again, the first notable difference between GR and the dissipative UG equations is that, in the latter, the CC is merely an integration constant. Also, in these equations, $Q$ is an arbitrary function that measures the energy diffusion, i.e., the nonconservation of the energy-momentum tensor in UG, and $\Pi=-3H\xi$ is the bulk viscous pressure in Eckart's theory, as we saw before.

In this work, we propose the following Ansätze for the diffusion function and the bulk viscosity:
\begin{equation}\label{diffusion}
    Q=\nu H^{2},
\end{equation}
\begin{equation}\label{bulkvicosity}
    \xi=\xi_{0}|Q|^{1/2}=\xi_{0}^{\nu}H,
\end{equation}
respectively, where $\nu$ is an arbitrary dimensionless parameter and $\xi_{0}^{\nu}\equiv \xi_{0}|\nu|^{1/2}$ is also a dimensionless parameter, with $\xi_{0}>0$ and, therefore, $\xi>0$ in order to be consistent with the second law of thermodynamics~\cite{Weinberg:1971mx}. The motivations for these choices are\footnote{In Section~\ref{sec:DifussionAndBulk}, we discuss some  physical arguments supporting the
hypothesis that cosmological dissipation naturally arises
from energy non-conservation via diffusion.}:

\begin{itemize}
\item[\textbf{i)}] As previously mentioned, at the classical level, UG is the simplest modification of GR that allows for the inclusion of non-conservative phenomena \cite{Josset:2016vrq, pellecchia2026darkenergygenesismodeling}. Nevertheless, the evolution equations in this framework must be completed by specifying a well-defined diffusion function $Q(t)$ to obtain well-defined dynamics. One can instead propose a phenomenological ansatz and assume dissipative effects within the dark sector as an effective response to the energy flux present in the modified Eq.~(\ref{UGconservation}).
\item[\textbf{ii)}] Considering the physical argument that an expanding universe is not expected to have a static vacuum energy density, several models have been proposed in the literature, such as the running vacuum models, in which the vacuum energy density has a certain time dependence. In particular, the most general time dependence of this energy density is given by the implicit expression $\rho_{\text{vac}}(H,\dot{H})=\Lambda+\nu H^{2}+\tilde{\nu}\dot{H}$ (where we identify $3c_{0}\to\Lambda$, $3\nu\to\nu$, and $3\tilde{\nu}\to\tilde{\nu}$ in~\cite{Sola:2015rra}), with $\nu$ and $\tilde{\nu}$ being dimensionless parameters motivated by perturbative results of Quantum Field Theory in a curved classical background~\cite{parker_toms_2009}. From Eqs.~\eqref{UGfluids}, \eqref{UGpressures}, and \eqref{diffusion}, we can define $p_{\text{vac}}=-\rho_{\text{vac}}$ with $\rho_{\text{vac}}\equiv\Lambda+Q=\Lambda+\nu H^{2}$, i.e., $\tilde{\nu}=0$, which is the most studied case in scenarios with variable vacuum energy density. Moreover, Eq.~\eqref{diffusion} is similar (but not identical) to the expression $Q\propto\rho$ studied in~\cite{Corral:2020lxt} in the standard UG framework (without dissipation), which shows competitive results in describing cosmological data from SNe~Ia and OHD.
\item[\textbf{iii)}] The parameterization $\xi\propto H$ is one of the most studied in the literature, in which the bulk viscosity is proportional to the expansion rate of the universe~\cite{Normann:2016jns}. A running vacuum model with bulk viscous DM in the GR framework was studied in Ref.~\cite{Sarath:2022odb,Cruz:2023dzn}. In the former, the parameterization $\xi=\xi_{0}H$ was used, while in the latter, the novel parameterization $\xi=\xi_{0}H\Omega_{m}^{s}$ was employed. The novelty of our work lies in the fact that the ``running'' is an apparent effect of the nonconservation of the energy-momentum tensor in UG~\cite{Corral:2020lxt}, and it is related to the bulk viscosity.
\item[\textbf{iv)}]  Our choice also has the following meaning in terms of the origin of the dissipative effect on the main fluid that fills the universe: since $\xi\propto |Q|^{1/2}$, the dissipative process that experiment the main fluid can be interpreted as a consequence of the energy flux $\dot{Q}$ that contributes to change the main fluid energy density through Eq.~(\ref{UGconservation}). Moreover, as Eq.(\ref{UGfluids}) is not modified by the inclusion of dissipation, the $Q$ term effectively influences the expansion rate, and therefore, it is more natural to assume the above interpretation for the Ansatz chosen for the dissipation. 

\end{itemize}
It is important to note that the bulk viscosity and the diffusion function are linked through the parameter $\nu$, for which, in the case of $\nu=0$, Eqs.~\eqref{UGfluids}–\eqref{UGpressures} are reduced to their GR counterparts, Eqs.~\eqref{GRfluids}–\eqref{GRconservation}. Nevertheless, the bulk viscosity has the independent parameter $\xi_{0}$, for which, in the case of $\xi_{0}=0$, we obtain the evolution equations for UG without dissipation. Therefore, we aim to study a dissipative UG model in which $\nu$ and $\xi_{0}$ are free parameters, considering the particular case $\xi_{0}=1$ ($\xi_{0}^{\nu}=|\nu|^{1/2}$). We also study the UG model obtained when $\xi_{0}=0$ ($\xi_{0}^{\nu}=0$).

From Eqs. \eqref{UGfluids}, \eqref{UGpressures}, \eqref{diffusion}, and \eqref{bulkvicosity}, we obtain the following first-order differential equation for the Hubble parameter
\begin{equation}\label{dHdt}
    \dot{H}=\frac{3H^{2}}{2}\left[\gamma\left(\frac{\nu}{3}-1\right)+\xi_{0}^{\nu} \right]+\frac{\gamma\Lambda}{2},
\end{equation}
where we have considered a barotropic EoS for the dissipative fluid of the form $p=(\gamma-1)\rho$, or $\omega\equiv\gamma-1$, and we assume a strictly expanding universe, i.e., $H>0$. From this differential equation, it is possible to obtain an analytic solution for the Hubble parameter as a function of the redshift $z$, considering the relation $1+z=a_{0}/a$ with $a_{0}$ the scale factor at the current time, given by
\begin{equation}\label{HdissipativeUG}
\begin{split}
    H^{2}(z)= H_{0}^{2} & \left\{\left[1-\frac{\gamma}{\alpha}\left(1-\Omega_{m,0}^{\nu}\right)\right]\left(1+z\right)^{(3-\nu)\alpha}\right. \\
    & \left. +\frac{\gamma}{\alpha}\left(1-\Omega_{m,0}^{\nu}\right)\right\},
\end{split}
\end{equation}
where $H_{0}$ is the Hubble parameter at the current time and
\begin{equation}\label{defofOmeganu}
    \Omega_{m,0}^{\nu}\equiv\frac{\Omega_{m,0}}{\left(1-\nu/3\right)},\;\; \Omega_{\Lambda,0}^{\nu}\equiv\frac{\Omega_{\Lambda,0}}{\left(1-\nu/3\right)},
\end{equation}
\begin{equation}\label{defofalpha}
    \alpha\equiv\gamma-\frac{3\xi_{0}^{\nu}}{\left(3-\nu\right)},
\end{equation}
with $\Omega_{m,0}\equiv\rho_{0}/3H_{0}^{2}$, $\Omega_{\Lambda,0}\equiv\Lambda/3H_{0}^{2}$, and $\Omega_{\Lambda,0}^{\nu}=1-\Omega_{m,0}^{\nu}$, this last constrain derived from Eqs. \eqref{UGfluids} and \eqref{diffusion}. Note that we have identified the dissipative fluid as the matter component of the universe. This solution was previously found in~\cite{Sarath:2022odb} with the identifications $\nu\to 3\nu$, $\xi_{0}^{\nu}\to\tilde{\xi}_{1}$, and for $\gamma=1$. Nevertheless, as we mentioned before, this solution has been obtained for a running vacuum model with dissipative DM in the context of GR. In this sense, we need to highlight that the novelty of the solution \eqref{HdissipativeUG} is that it is obtained in the context of UG for a particular election of the diffusion function, which leads to the same term as the one used in~\cite{Sarath:2022odb} for the running vacuum. Also, even though we consider the same parameterization for the bulk viscosity of the form $\xi\propto H$, in our work, this expression is related to the diffusion function, making our solution different from the one obtained in~\cite{Sarath:2022odb}. Therefore, the solution \eqref{HdissipativeUG} is different in nature, and the relationship between the diffusion function and the bulk viscosity opens up new possibilities to make this solution observationally competitive concerning the $\Lambda$CDM model. 

Note that Eq. \eqref{HdissipativeUG} has the following four particular solutions:
\begin{itemize}
    \item[\textbf{i)}] Consider that in UG $\Lambda$ is an integration constant, then it can, in principle, take the values $\Lambda>0$, $\Lambda=0$, and $\Lambda<0$. In this sense, solution \eqref{HdissipativeUG} is valid for the cases $\Lambda>0$ and $\Lambda<0$, while, for $\Lambda=0$, the solution becomes
    \begin{equation}\label{Honlymatter}
        H^{2}(z)=H^{2}_{0}\left(1+z\right)^{\left[\gamma(3-\nu)-3\xi^{\nu}_{0}\right]},
    \end{equation}
    which implies that $\Omega_{m,0}^{\nu}=1$ or, equivalently, $\Omega_{m,0}=(3-\nu)/3$.
    \item[\textbf{ii)}] From Eq. \eqref{defofOmeganu}, it is clear that $\nu\neq 3$. This is because $\nu=3$ leads to the solution
    \begin{equation}\label{HDeSitter}
        \rho_{m}=-\Lambda,
    \end{equation}
    which is easily obtained from Eqs. \eqref{UGfluids} and \eqref{diffusion}. Therefore, this solution is discarded because it represents a universe with a negative energy density for matter in the standard case in which $\Lambda>0$.
    \item[\textbf{iii)}] For the particular case in which the matter fluid does not experience dissipative processes ($\xi_{0}=0$), the solution \eqref{HdissipativeUG} becomes
    \begin{equation}\label{HUG}
        H^{2}(z)=H_{0}^{2}\left[\Omega_{m,0}^{\nu}\left(1+z\right)^{(3-\nu)\gamma}+1-\Omega_{m,0}^{\nu}\right],
    \end{equation}
    solution that is similar to the previous one found in Ref. \cite{Corral:2020lxt}. Nevertheless, the difference between solutions is that, here, the diffusion function has the form $Q\propto H^{2}$, while in Ref. \cite{Corral:2020lxt}, the diffusion function has the form $Q\propto\rho$, which are different in nature.
    \item[\textbf{iv)}] Finally, for $\nu=0$ and $\gamma=1$ (CDM), the solution \eqref{HdissipativeUG} becomes
    \begin{equation}\label{HLCDM}
        H^{2}(z)=H_{0}^{2}\left[\Omega_{m,0}\left(1+z\right)^{3}+1-\Omega_{m,0}\right],
    \end{equation}
    which is the Hubble parameter for the standard $\Lambda$CDM model at late times; a result expected from our previous analysis.
\end{itemize}

Since we are interested in the capability of this model in describing the late-time cosmological observational data, from now on, we consider only the cases in which $\Lambda>0$ and we fix the value $\gamma=1$, which represents a universe filled with dust matter. Also, we assume that all matter sector experiences dissipative processes and diffusion in the same way. This means that bayronic and DM evolve equally and, therefore, $\Omega_{m,0}=\Omega_{b,0}+\Omega_{DM,0}$, where $\Omega_{b,0}$ and $\Omega_{DM,0}$ are the density parameter of baryon and DM, respectively.

%%%%%%%%%%%%%%%%%%%%%%%%%%%%%%%%%%%%%%%%%%%%%%%%%%%%%%%%%
\section{\label{sec:Constraints}Cosmological Constraints}
%%%%%%%%%%%%%%%%%%%%%%%%%%%%%%%%%%%%%%%%%%%%%%%%%%%%%%%%%
In this section, we shall constrain the free parameters of the UG model under study. To achieve this, we compute the best-fit parameters at the $1\sigma\,(68.3\%)$ confidence level (CL) using the affine-invariant Markov chain Monte Carlo (MCMC) method \cite{Goodman_Ensemble_2010}, implemented in the pure-Python code \textit{emcee} \cite{Foreman-Mackey:2012any}. In this procedure, we consider 100 chains or ``walkers", using the autocorrelation time ($\tau_{\text{corr}}$), provided by the \textit{emcee} module, as a convergence test. The value of $\tau_{\text{corr}}$ for each free parameter is computed at every $50$ steps. Then, if the current step exceeds $50\tau_{\text{corr}}$ and the value of $\tau_{\text{corr}}$ changes by less than $1\%$, we conclude that the chains have converged and the constraint is stopped. This convergence test is complemented by computing the mean acceptance fraction of the chains, which must have a value between $0.2$ and $0.5$ \citep{Foreman-Mackey:2012any} and can be modified by the stretch move provided by the \textit{emcee} module. The first $5\tau_{\text{corr}}$ are discarded as ``burn-in" steps, and the statistical analysis is performed on a chain that is flattened and thinned by $\tau_{\text{corr}}/2$. 

In this MCMC analysis, the posterior distribution $P(\boldsymbol{\theta},X)$, i.e., the probability of the free parameters $\boldsymbol{\theta}$ given the observational data $\boldsymbol{X}$, is computed via Bayes' Theorem
\begin{equation}\label{Bayes}
    P(\boldsymbol{\theta,X})=\frac{P(\boldsymbol{X,\theta})P(\boldsymbol{\theta})}{P(\boldsymbol{X})},
\end{equation}
where $P(\boldsymbol{\theta})$ is the prior distribution and corresponds to the previous physical evidence about the free parameters space $\boldsymbol{\theta}$, $P(\boldsymbol{X})$ is the prior predictive probability which is approximated by the MCMC procedure, and $P(\boldsymbol{X,\theta})$ is the likelihood function, for which we consider the following Gaussian distribution:
\begin{equation}\label{likelihood}
\mathcal{L}_{I}\propto\exp{\left(-\frac{\chi_{\text{I}}^{2}}{2}\right)},
\end{equation}
where $\chi_{\text{I}}^{2}$ is the merit function and I stands for each data set consider for the constraint, including the joint analysis in which $\chi^{2}_{\text{joint}}=\sum_{I}\chi_{\text{I}}^{2}$. Since the best-fit parameters minimize the $\chi^{2}$ function, its minimum value can be used as an indicator of the goodness of fit: the smaller the value of $\chi_{\text{min}}^{2}$ is, the better the fit.

In principle, the value of $\chi_{\text{min}}^{2}$ obtained for the best-fit parameters
can be reduced by adding free parameters to the model under study, potentially resulting in overfitting. Hence, we compute the Bayesian Information Criterion (BIC) \cite{Schwarz:1978tpv} to compare the goodness of the fit statistically. This criterion adds a penalization to the value of $\chi_{\text{min}}^{2}$, that depends on the total number of free parameters of the model ($\theta_{N}$), according to the expression
\begin{equation}
    BIC=\theta_{N}\ln{(n)}+\chi_{\text{min}}^{2},
\end{equation}
where $n$ is the total number of data points in the corresponding data sample. Thus, when two different models are compared, the one most favored by the observations statistically corresponds to the one with the smallest value of BIC. In general, a difference of $2-6$ in BIC is evidence against the model with the higher BIC, a difference of $6-10$ is strong evidence, and a difference of $>10$ is very strong evidence.

In the following subsections, we provide a brief overview of the merit function for each data set included in the cosmological constraints.

%%%%%%%%%%%%%%%%%%%%%%%%%%%%%%%%%%%%%%%%%%%%%%%%%%%
\subsection{\label{subsec:SNeIa}Type Ia supernovae}
%%%%%%%%%%%%%%%%%%%%%%%%%%%%%%%%%%%%%%%%%%%%%%%%%%%
Supernovae (SNe) are highly energetic explosions of some stars and play an important role in the fields of astrophysics and cosmology because they have been used as cosmic distance indicators. In particular, Type Ia Supernovae (SNe Ia) are considered standard candles to measure the geometry and the late-time dynamics of the Universe \cite{Liu:2023qmw}. For SNe Ia data, we consider the Pantheon+ sample \cite{Brout:2022vxf}, which consists of 1701 data points in the redshift range $0.001\leq z\leq 2.26$, whose respective merit function can be conveniently constructed in matrix notation (denoted by bold symbols) as
\begin{equation}\label{meritSNe}
\chi_{\text{SNe}}^{2}=\mathbf{\Delta D}(z,\boldsymbol{\theta},M)^{\dagger}\mathbf{C}^{-1}\mathbf{\Delta D}(z,\boldsymbol{\theta},M),
\end{equation}
where $\left[\mathbf{\Delta D}(z,\boldsymbol{\theta},M)\right]_{i}= m_{B,i}-M-\mu_{th}(z_{i},\boldsymbol{\theta})$ and $\mathbf{C}=\mathbf{C}_{\text{stat}}+\mathbf{C}_{\text{sys}}$, with $\mathbf{C}$ being the total uncertainty covariance matrix. The matrices $\mathbf{C}_{\text{stat}}$ and $\mathbf{C}_{\text{sys}}$ account for the statistical and systematic uncertainties, respectively. The quantity $\mu_{i}=m_{B, i}-M$ corresponds to the observational distance modulus of the Pantheon+ sample, which is obtained by a modified version of Trip's formula \cite{Tripp:1997wt} and the BBC (BEAMS with Bias Corrections) approach \cite{Kessler:2016uwi}. In contrast, $m_{B, i}$ is the corrected apparent B-band magnitude of a fiducial SNe Ia at redshift $z_{i}$, and $M$ is the fiducial magnitude of a SNe Ia, which must be jointly estimated with the free parameters of the model under study. On the other hand, the theoretical  distance modulus for a spatially flat FLRW spacetime is given by
\begin{equation}\label{theoreticaldistance}
\mu_{th}(z_{i},\boldsymbol{\theta})=5\log_{10}{\left[\frac{d_{L}(z_{i},\boldsymbol{\theta})}{\text{Mpc}}\right]}+25,
\end{equation}
with $d_{L}(z_{i},\boldsymbol{\theta})$ the  luminosity distance given by
\begin{equation}\label{luminosity}
d_{L}(z_{i},\boldsymbol{\theta})=c(1+z_{i})\int_{0}^{z_{i}}{\frac{dz'}{H_{th}(z',\boldsymbol{\theta})}},
\end{equation}
where $c$ is the speed of light given in units of $\text{km/s}$.

In principle, there is a degeneracy between $M$ and $H_{0}$. Hence, to constrain $H_{0}$ using SNe Ia data alone, it is necessary to include the SH0ES (Supernovae and $H_{0}$ for the Equation of State of the dark energy program) Cepheid host distance anchors, with a merit function of the form
\begin{equation}\label{Cepheidmerit}
\chi^{2}_{\text{Cepheid}}=\mathbf{\Delta D}_{\text{Cepheid}}\left(M\right)^{\dagger}\textbf{C}^{-1}\mathbf{\Delta D}_{\text{Cepheid}}\left(M\right),
\end{equation}
where 
$\left[\mathbf{\Delta D}_{\text{Cepheid}}\left(M\right)\right]_{i}=\mu_{i}\left(M\right)-\mu_{i}^{\text{Cepheid}}$, where $\mu_{i}^{\text{Cepheid}}$ is the Cepheid calibrated host-galaxy distance obtained by SH0ES \cite{Riess:2021jrx}. Thus, we use the Cepheid distances as the ``theory model'' to calibrate $M$, considering that the difference $\mu_{i}\left(M\right)-\mu_{i}^{\text{Cepheid}}$ is sensitive to $M$ and largely insensitive to other parameters of the cosmological model. Therefore, taking into account that the total uncertainty covariance matrix for Cepheid is contained in the total uncertainty covariance matrix $\mathbf{C}$, we define the merit function for the SNe Ia data as
\begin{equation}\label{SNemeritfull}
\chi_{\text{SNe}}^{2}=\mathbf{\Delta D'}(z,\boldsymbol{\theta},M)^{\dagger}\mathbf{C}^{-1}\mathbf{\Delta D'}(z,\boldsymbol{\theta},M),
\end{equation}
where
\begin{equation}\label{SNeresidual}
\Delta\mathbf{D'}_{i}=\left\{\begin{array}{ll}
m_{B,i}-M-\mu_{i}^{\text{Cepheid}} & i\in\text{Cepheid host} \\
\\ m_{B,i}-M-\mu_{th}(z_{i},\boldsymbol{\theta}) & \text{otherwise}
\end{array}
\right..
\end{equation}

%%%%%%%%%%%%%%%%%%%%%%%%%%%%%%%%%%%%%%%%%%%%%%%%%
\subsection{\label{subsec:CC}Cosmic Chronometers}
%%%%%%%%%%%%%%%%%%%%%%%%%%%%%%%%%%%%%%%%%%%%%%%%%
Even though SNe Ia data provide consistent evidence about the existence of a transition epoch in cosmic history where the expansion rate of the Universe changes, it is important to highlight that this conclusion is obtained in a model-dependent way \cite{Moresco:2016mzx}. The study of the expansion rate of the Universe in a model-independent way can be carried out through measurements of the Hubble parameter compiled in the Cosmic Chronometers (CC) data sample.
In this paper, we consider the data set from Ref. \cite{Capozziello:2017nbu}, which comprises 31 Hubble parameter data points spanning the redshift range $0.0708\leq z\leq 1.965$. In this case, the merit function can be directly constructed as follows:
\begin{equation}\label{meritCC}
    \chi_{CC}^{2}=\sum_{i=1}^{31}{\left[\frac{H_{i}-H_{th}(z_{i},\boldsymbol{\theta)}}{\sigma_{H,i}}\right]^{2}},
\end{equation}
where $H_{i}$ represents the observational Hubble parameter data at redshift $z_{i}$, with an associated uncertainty $\sigma_{H,i}$, all provided by the CC sample, and $H_{\text{th}}$ denotes the theoretical Hubble parameter value at the same redshift. It is important to mention that these Hubble parameter data points are derived using the differential age method, a model-independent approach \cite{Jimenez:2001gg}. 

%%%%%%%%%%%%%%%%%%%%%%%%%%%%%%%%%%%%%%%%%%%%%%%%%%%%%%%%%%
\subsection{\label{subsec:BAO}Baryon acoustic oscillation}
%%%%%%%%%%%%%%%%%%%%%%%%%%%%%%%%%%%%%%%%%%%%%%%%%%%%%%%%%%
Measurements of baryon acoustic oscillations (BAO) provide a powerful tool for studying the expansion history of the Universe. The method is based on a characteristic scale that was imprinted on matter clustering by pressure waves propagating through the coupled photon-baryon fluid of the pre-recombination era \cite{Eisenstein:1998tu,Blake:2003rh,Seo:2003pu}. In other words, BAO exploits the enhancement of clustering at the scale of the pre-recombination sound horizon \cite{Weinberg:2013agg}, given by
\begin{equation}\label{sh}
    r_{d}=\int_{z_{d}}^{\infty}{\frac{C_{s}(z)}{H(z)}}dz,
\end{equation}
where $C_{s}(z)$ is the speed of sound in the photon-baryon fluid and $z_{d}\approx 1060$ \cite{Planck:2018vyg} is the redshift at which acoustic
waves stall because photons no longer ``drag'' the baryons. Assuming a standard cosmological model in the pre-recombination epoch, in which the universe is composed of baryons, CDM, photons, and other relativistic species (which does not necessarily imply $\Lambda$CDM cosmology) \cite{Brieden:2022heh}, the sound horizon above can be written as
\begin{eqnarray}\label{shstandard}
    r_{d}&&=147.05\,Mpc\left(\frac{\Omega_{B,0}h^{2}}{0.02236}\right)^{-0.13}\notag \\
    &&\times \left[\frac{(\Omega_{B,0}+\Omega_{DM,0})h^{2}}{0.1432}\right]^{-0.23}\left(\frac{N_{eff}}{3.04}\right)^{-0.1},
\end{eqnarray}
where $H_{0}=100\frac{km/s}{Mpc}h$, with $h$ being the reduced Hubble parameter, and $N_{eff}=3.044$ represents the effective number of relativistic degrees of freedom for three neutrinos at $z>z_{d}$. The above equation is scaled to the best-fit values obtained from the Planck collaboration \cite{Planck:2018vyg}, and $\Omega_{B,0}$ and $\Omega_{DM,0}$ represent the density parameters of baryons and DM, respectively. 

Concerning the BAO measurements, the BAO scale in the transverse direction for a spatially flat FLRW spacetime is given by $d_{L}(z_{i},\boldsymbol{\theta})=(1+z_{i})d_{M}(z_{i},\boldsymbol{\theta})$, or equivalently, according to Eq. \eqref{luminosity},
\begin{equation}\label{BAODM}
    d_{M}(z_{i},\boldsymbol{\theta})=c\int_{0}^{z_{i}}{\frac{dz'}{H_{th}(z',\boldsymbol{\theta})}dz'},
\end{equation}
which allows for constraining the transverse comoving distance. On the other hand, the BAO measurement in the line-of-sight direction is given by
\begin{equation}\label{BAODH}
    d_{H}(z_{i},\boldsymbol{\theta})=\frac{c}{H_{th}(z_{i},\boldsymbol{\theta})},
\end{equation}
which allows us to constrain the expansion rate of the universe, $H(z)$.

In this paper, we consider the DESI DR2 sample (see Table IV of Ref. \cite{DESI:2025zgx}), whose respective inferred BAO distances are given relative to the sound horizon, i.e., $d_{M,i}/r_{d}$ and $d_{H,i}/r_{d}$, except for one data point, which corresponds to the isotropic BAO distance, given by
\begin{equation}\label{BAODV}
    d_{V}(z_{i},\boldsymbol{\theta})=\left[z_{i}d_{M}^{2}(z_{i},\boldsymbol{\theta})d_{H}(z_{i},\boldsymbol{\theta})\right]^{1/3}.
\end{equation}
Therefore, the merit function for the BAO measurements is constructed as
\begin{equation}\label{meritBAO}
    \chi_{\text{BAO}}^{2}=\sum_{i=1}^{13} 
{\left[\frac{\Delta d_{i}(z_{i},\boldsymbol{\theta})}{\sigma_{D,i}}\right]^{2}},
\end{equation}
where $\sigma_{D,i}$ is the associated uncertainty and
\begin{equation}\label{BAOdata}
\Delta d_{i}(z_{i},\boldsymbol{\theta})=\left\{\begin{array}{lll}
\left[d_{M,i}-d_{M}(z_{i},\boldsymbol{\theta})\right]/r_{d}, & i\in A \\
\\ \left[d_{H,i}-d_{H}(z_{i},\boldsymbol{\theta})\right]/r_{d}, & i\in A \\
\\ \left[d_{V,i}-d_{V}(z_{i},\boldsymbol{\theta})\right]/r_{d}, & i\in B
\end{array}
\right.,
\end{equation}
with $A=\{\text{LRG1},\text{LRG2},\text{LRG3}+\text{ELG1},\text{ELG2},\text{QSO},\allowbreak\text{Lya}\}$ and $B=\{\text{BGS}\}$.

For BAO measurements, there is a degeneracy between $H_{0}$ and $r_{d}$, which can be broken by knowing the value of $\Omega_{B,0}h^{2}$. Therefore, following Ref. \cite{DESI:2025zgx}, we consider a Gaussian prior on $\Omega_{B,0}h^{2}$ centered at $0.02218\pm 0.00055$, referred to as the Big Bang Nucleosynthesis prior.

%%%%%%%%%%%%%%%%%%%%%%%%%%%%%%%%%%%%%%%%%%%%%%%%%%%%%%%%
\subsection{\label{subsec:lensing}Gravitational Lensing}
%%%%%%%%%%%%%%%%%%%%%%%%%%%%%%%%%%%%%%%%%%%%%%%%%%%%%%%%
Multiple images are produced when a background object (the source) is lensed due to the gravitational force of a massive body (the lens). Therefore, the light rays emitted from the source will take different paths through spacetime at different image positions and arrive at the observer at different times. In this sense, the time delay of two different images $k$ and $l$ depends on the mass distribution along the line-of-sight of the lensing object, which can be calculated as follows:
\begin{equation}\label{lensing}
    \Delta t_{kl}=\frac{d_{\Delta t}}{c}\left[\frac{\left(\phi_{k}-\beta\right)^{2}}{2}-\psi(\phi_{k})-\frac{\left(\phi_{l}-\beta\right)^{2}}{2}+\psi(\phi_{l})\right],
\end{equation}
where $\phi_{k}$ and $\phi_{l}$ are the angular positions of the images, $\beta$ is the angular positions of the source, $\psi(\phi_{k})$ and $\psi(\phi_{l})$ are the lens potentials at the image positions, and $d_{\Delta t}$ is the ``time-delay distance'', which is theoretically given by the expression \cite{Treu:2016ljm}
\begin{equation}\label{time-delay}
    d_{\Delta t}^{th}(\mathbf{z},\boldsymbol{\theta})=\left(1+z_{l}\right)\frac{d_{A,l}(z_{l},\boldsymbol{\theta})d_{A,s}(z_{s},\boldsymbol{\theta})}{d_{A,ls}(z_{ls},\boldsymbol{\theta})},
\end{equation}
where the subscripts $l$, $s$, and $ls$ stand for the lens, the source, and between the lens and the source, respectively; $\mathbf{z}=(z_{l},z_{s},z_{ls})$ and $d_{A,j}$ is the angular diameter distance, which can be written in terms of the luminosity distance \eqref{luminosity} as $d_{L}(z_{j},\boldsymbol{\theta})=d_{A,j}(1+z_{j})^{2}$, or
\begin{equation}\label{angulardistance}
    d_{A,j}(z_{j},\boldsymbol{\theta})=\frac{c}{(1+z_{j})}\int_{0}^{z_{j}}{\frac{dz'}{H_{th}(z',\boldsymbol{\theta})}}.
\end{equation}

In this paper, we consider the Gravitational Lensing (GL) compilation provided by the H0LiCOW collaboration \cite{Wong:2019kwg}, which consists of six lensed quasars: B1608+656 \cite{Jee:2019hah}, SDSS 1206+4332 \cite{Birrer:2018vtm}, WFI2033-4723 \cite{Rusu:2019xrq}, RXJ1131-1231, HE 0435-1223, and PG 1115-080 \cite{Chen:2019ejq}; whose respective merit function can be written as
\begin{equation}\label{H0LiCOWmerit}
    \chi^{2}_{\text{GL}}=\sum_{i=1}^{10}{\left[\frac{d_{\Delta t,i}-d_{\Delta t}^{th}(\mathbf{z}_{i},\boldsymbol{\theta})}{\sigma_{d_{\Delta t},i}}\right]^{2}},
\end{equation}
where $d_{\Delta t, i}$ is the observational time-delay distance of the lensed quasar at redshift $\mathbf{z}_{i}=(z_{l, i};z_{s, i};z_{ls, i})$ with an associated uncertainty $\sigma_{d_{\Delta t}, i}$ (for more details, see Ref. \cite{Wong:2019kwg}). It is important to note that, for $z\to 0$, the angular diameter distance \eqref{angulardistance} tends to $d_{A}\to cz/H_{0}$ and, therefore, the gravitational lensing data of the H0LiCOW collaboration is sensitive to $H_{0}$, with a weak dependency on other cosmological parameters.

%%%%%%%%%%%%%%%%%%%%%%%%%%%%%%%%%%%%%%%%%%%%%%%%%
\subsection{\label{subsec:BHS}Black Hole Shadows}
%%%%%%%%%%%%%%%%%%%%%%%%%%%%%%%%%%%%%%%%%%%%%%%%%
The Black Hole Shadows (BHS) data are of interest for studying our local universe since their dynamics are quite simple and can be seen as standard rulers if the angular size redshift $\alpha$, the relation between the size of the shadow and the mass of the supermassive black hole that produces it, is established \cite{Escamilla-Rivera:2022mkc}. In this paper, we are interested in two measurements: the first one was made on the M87* supermassive black hole by The Event Horizon Telescope Collaboration \cite{EventHorizonTelescope:2019dse} (the first detection of a BHS), and the second one corresponds to the detection of Sagittarius A* (Sgr A*) \cite{EventHorizonTelescope:2022wkp}.

Light rays curve around its event horizon in a black hole (BH), creating a ring with a black spot at its center, the so-called shadow of the BH. In this sense, the angular radius of the BHS for a Schwarzschild (SH) BH at redshift $z_{i}$ can be written as
\begin{equation}\label{angularradius}
    \alpha_{SH}\left(z_{i},\boldsymbol{\theta}\right)=\frac{3\sqrt{3}m}{d_{A}(z_{i},\boldsymbol{\theta})},
\end{equation}
where $d_{A}(z_{i},\boldsymbol{\theta})$ is given by Eq. \eqref{angulardistance} (note that the sub-index $j$ is not necessary in this case) and $m=GM_{BH}/c^{2}$ is the mass parameter of the BH, with $M_{BH}$ the mass of the BH in solar masses units and $G$ the gravitational constant. It is common to write Eq. \eqref{angularradius} in terms of the shadow radius $\alpha_{SH}(z_{i},\boldsymbol{\theta})=R_{SH}/d_{A}(z_{i},\boldsymbol{\theta})$, where $R_{SH}=3\sqrt{3}GM_{BH}/c^{2}$ (the speed of light is given in units of $\text{m/s}$ in this case). Therefore, the merit function for the BHS data can be constructed as
\begin{equation}\label{BHSmerit}
    \chi^{2}_{BHS}=\sum_{i=1}^{2}{\left[\frac{\alpha_{i}-\alpha_{SH}(z_{i},\boldsymbol{\theta})}{\sigma_{\alpha,i}}\right]^{2}},
\end{equation}
where $\alpha_{i}$ is the observational angular radius of the BHS at redshift $z_{i}$ with an associated uncertainty $\sigma_{\alpha,i}$. It is important to note that for $z\to 0$ the angular radius \eqref{angularradius} tends to $\alpha_{SH}\to R_{SH}H_{0}/cz$ and, therefore, similar to the GL data, the BHS data is sensitive to $H_{0}$, with a weak dependency on other cosmological parameters. On the other hand, Eq. \eqref{angularradius} is divided by a factor of $1.496\times 10^{11}$ to obtain $\alpha_{SH}$ in units of $\mu as$.

%%%%%%%%%%%%%%%%%%%%%%%%%%%%%%%%%%%%%%%%%%%%%%%%%%%%%%%%%%%%%
\subsection{\label{subsec:priors}Theoretical Hubble parameter and priors}
%%%%%%%%%%%%%%%%%%%%%%%%%%%%%%%%%%%%%%%%%%%%%%%%%%%%%%%%%%%%%
Regardless of the dataset considered in the constraints, we need to define the theoretical Hubble parameter for the cosmological model under study. In our case, we constrain the cosmological model given by Eq. \eqref{HdissipativeUG}, whose respective free parameters are $\boldsymbol{\theta}=\{h,\Omega_{b,0},\Omega_{DM,0},\nu,\xi_{0}^{\nu}\}$, where we have fixed the value $\gamma=1$, as previously mentioned. We should emphasize that we are also interested in the cases where $\xi_{0}^{\nu}=|\nu|^{1/2}$ ($\xi_{0}=1$) and in the UG model without dissipation, in which $\xi_{0}^{\nu}=0$ ($\xi_{0}=0$), as given by Eq. \eqref{HUG}. Therefore, we consider the following flat (F) and Gaussian (G) priors for our MCMC analysis:
\begin{itemize}
    \item $h\in F(0.55,0.85)$,
    \item $\Omega_{B,0}h^{2}\in G(0.02218,0.00055)$,
    \item $\Omega_{DM,0}\in F(0,1)$,
    \item $\nu\in F(-3,3)$,
    \item $\xi_{0}^{\nu}\in F(0,\sqrt{3})$,
    \item $M\in F(-20,-18)$,
\end{itemize}
\noindent where $H_{0}=100\frac{km/s}{Mpc}h$ and $M$ is the nuisance parameter of the SNe Ia data. Furthermore, we also constrain the $\Lambda$CDM model for comparison, whose theoretical Hubble parameter at late times is given by Eq. \eqref{HLCDM}. Consequently, its free parameters are $\boldsymbol{\theta}=(h,\Omega_{b,0},\Omega_{DM,0})$, since $\Omega_{m,0}=\Omega_{b,0}+\Omega_{DM,0}$, for which we use the same priors as in the UG model.

It is important to note that, to avoid a complex Hubble parameter in our MCMC analysis in the $\xi_{0}^{\nu}\neq 0$ cases, we also impose the following condition as a constraint:
\begin{equation}\label{condition}
    1-\frac{(1-\Omega_{m,0}^{\nu})}{\alpha}>0.
\end{equation}

%%%%%%%%%%%%%%%%%%%%%%%%%%%%%%%%%%%%%%%%%%%%%%%%%%%%
\section{\label{sec:Results}Results and Discussions}
\begin{figure}[ht]
    \centering
    \includegraphics[width=\columnwidth]{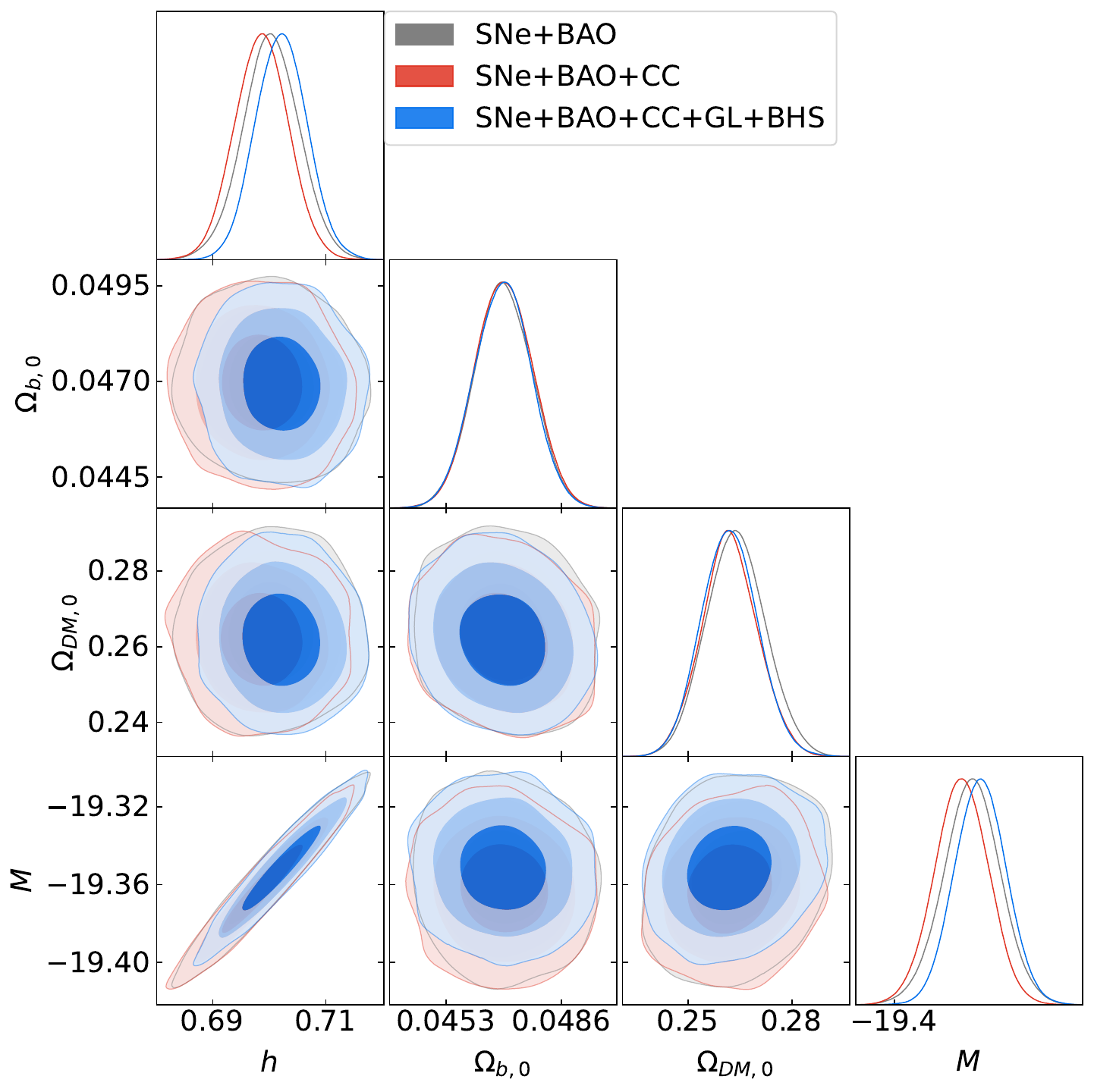}
    \caption{The posterior 1D distributions and joint marginalized regions for the free parameter space of the $\Lambda$CDM model, obtained via the MCMC analysis described
    in Section \ref{sec:Constraints}. The admissible joint regions correspond to the $1\sigma$, $2\sigma$, and $3\sigma$ CL, respectively. The best-fit values for each free parameter are shown in Table \ref{tab:best-fits}.}
    \label{fig:TriangleLCDM}
\end{figure}

\begin{figure}[ht]
    \centering
    \includegraphics[width=\columnwidth]{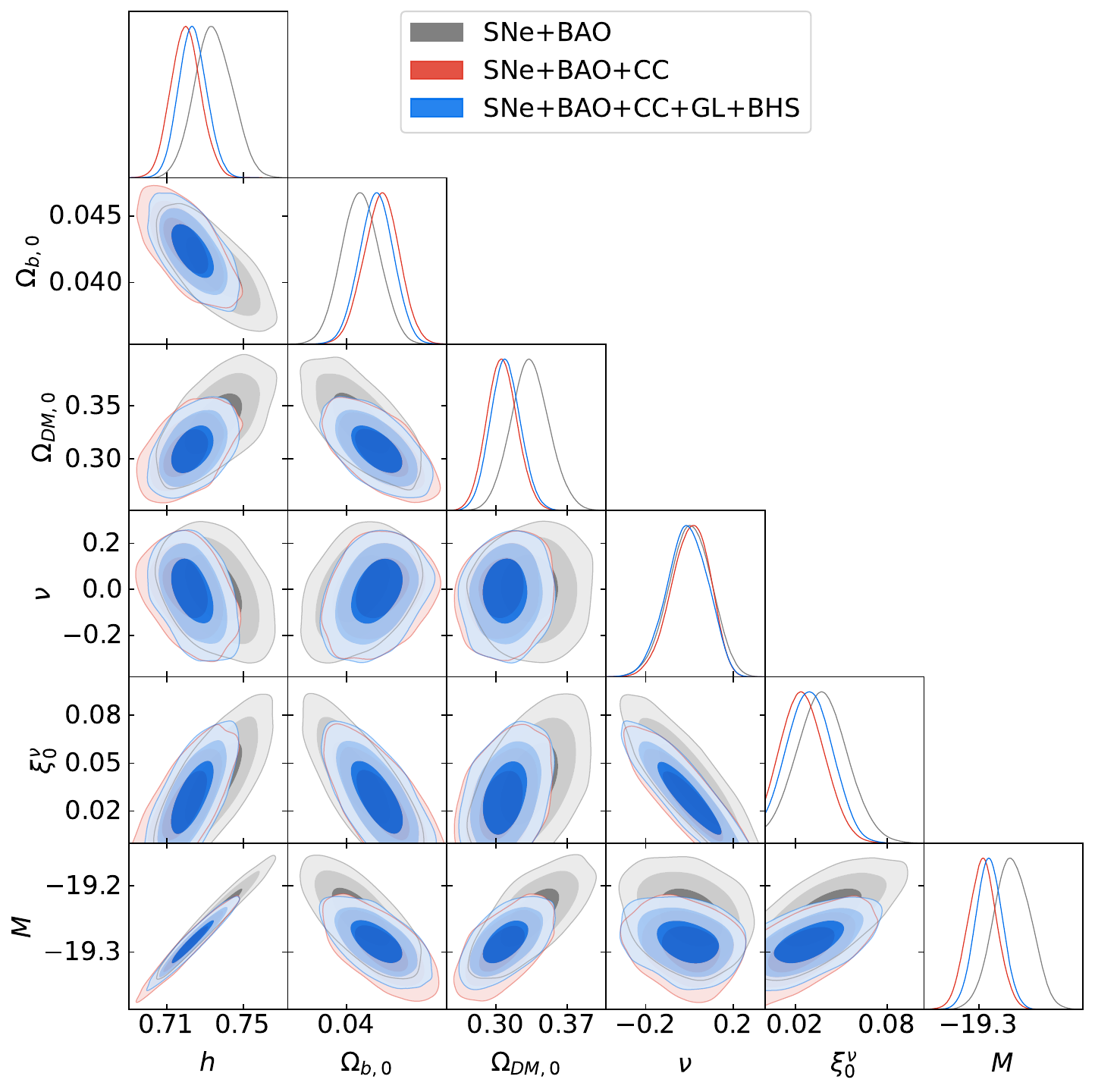}
    \caption{The posterior 1D distributions and joint marginalized regions for the free parameter space of the dissipative UG model, obtained via the MCMC analysis described
    in Section \ref{sec:Constraints}. The admissible joint regions correspond to the $1\sigma$, $2\sigma$, and $3\sigma$ CL, respectively. The best-fit values for each free parameter are shown in Table \ref{tab:best-fits}.}
    \label{fig:TriangleDissipativeUG}
\end{figure}

\begin{figure}[ht]
    \centering
    \includegraphics[width=\columnwidth]{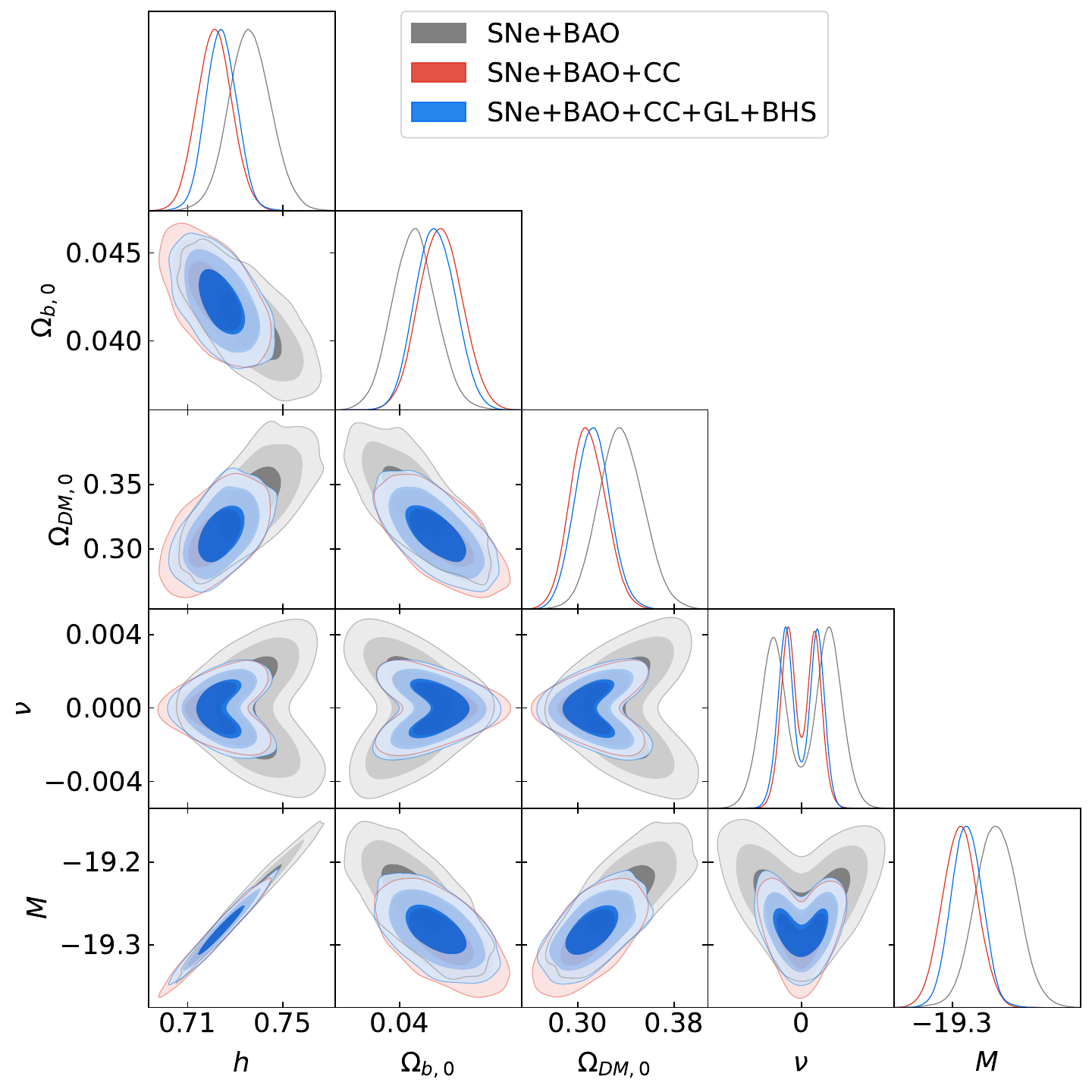}
    \caption{The posterior 1D distributions and joint marginalized regions for the free parameter space of the dissipative UG model with $\xi_{0}=1$ ($\xi_{0}^{\nu}=|\nu|^{1/2}$), obtained via the MCMC analysis described
    in Section \ref{sec:Constraints}. The admissible joint regions correspond to the $1\sigma$, $2\sigma$, and $3\sigma$ CL, respectively. The best-fit values for each free parameter are shown in Table \ref{tab:best-fits}.}
    \label{fig:TriangleParticularUG}
\end{figure}

\begin{figure}[ht]
    \centering
    \includegraphics[width=\columnwidth]{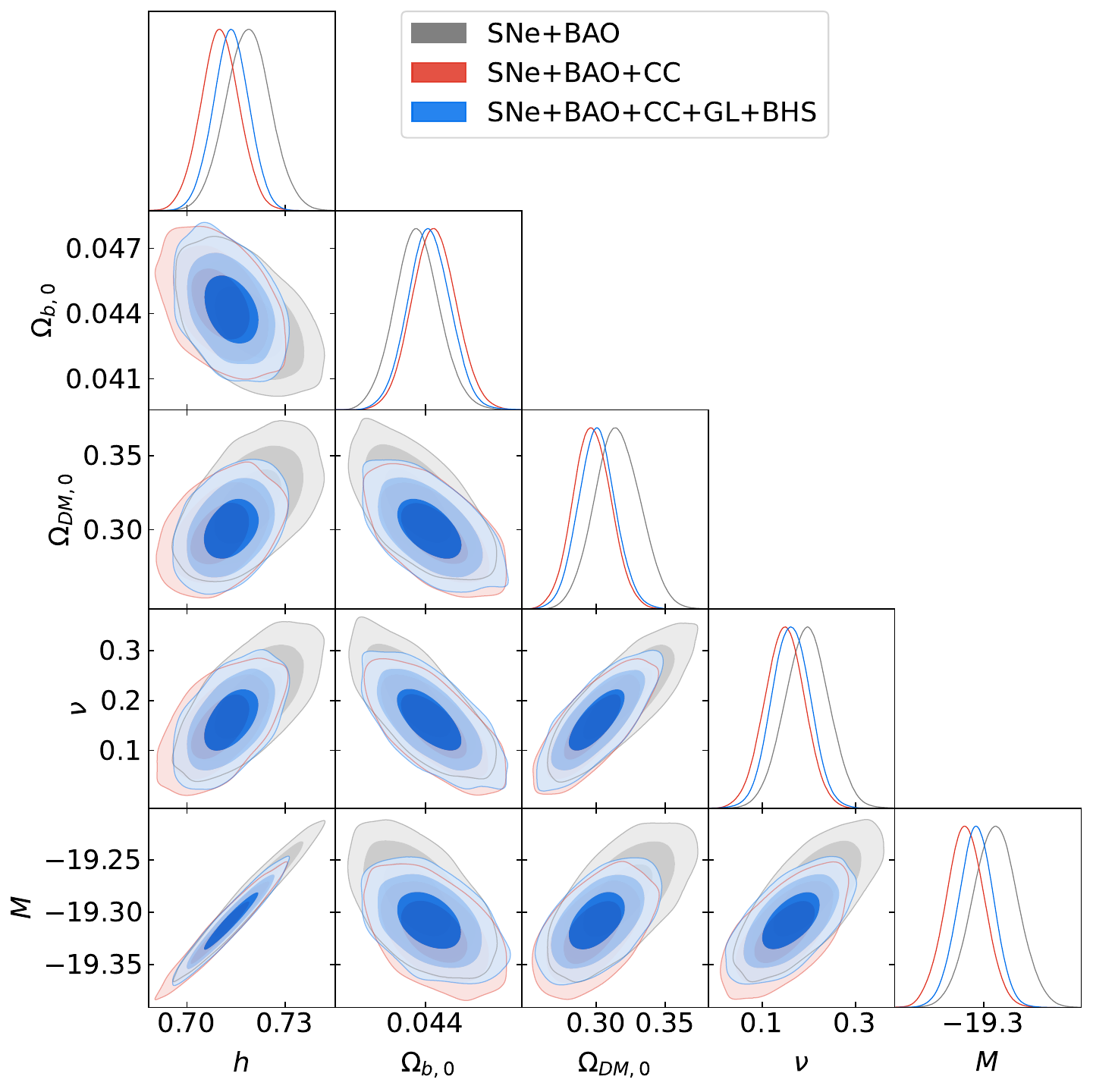}
    \caption{The posterior 1D distributions and joint marginalized regions for the free parameter space of the UG model without dissipation, obtained via the MCMC analysis described
    in Section \ref{sec:Constraints}. The admissible joint regions correspond to the $1\sigma$, $2\sigma$, and $3\sigma$ CL, respectively. The best-fit values for each free parameter are shown in Table \ref{tab:best-fits}.}
    \label{fig:TriangleUG}
\end{figure}

%%%%%%%%%%%%%%%%%%%%%%%%%%%%%%%%%%%%%%%%%%%%%%%%%%%%
In Table \ref{tab:best-fits}, we present the best-fit values at $1\sigma$ CL and the goodness-of-fit criteria for the free parameters of the $\Lambda$CDM, dissipative UG, and UG without dissipation models, including the particular dissipative UG model case where $\xi_{0}=1$ ($\xi_{0}^{\nu}=|\nu|^{1/2}$). In Table \ref{tab:deltagoodness}, we show the differences $\Delta\chi_{\text{min}}^{2}=\left(\chi_{\text{min}}^{\Lambda\text{CDM}}\right)^{2}-\left(\chi_{\text{min}}^{\text{model}}\right)^{2}$ and $\Delta\text{BIC}=\text{BIC}_{\Lambda\text{CDM}}-\text{BIC}_{\text{model}}$ between the $\Lambda$CDM and the UG models. These differences aim to simplify the assessment of model preference; by definition, a positive value indicates that the data favor the model under study, whereas a negative value implies evidence in favor of the $\Lambda$CDM model. In Figs. \ref{fig:TriangleLCDM}, \ref{fig:TriangleDissipativeUG}, \ref{fig:TriangleParticularUG}, and \ref{fig:TriangleUG}, we depict the 1D posterior distributions and joint marginalized regions at $1\sigma$, $2\sigma(95.5\%)$, and $3\sigma(99.7\%)$ CL, for the free parameter spaces of $\Lambda$CDM, disspative UG, dissipative UG with $\xi_{0}=1$, and UG whithout dissipation models, respectively. These results were obtained using the MCMC analysis described in Section \ref{sec:Constraints}.

\begin{table*}[ht]
    \centering
    \setlength{\tabcolsep}{1pt}
    \begin{tabularx}{\textwidth}{p{2.3cm}YYYYYYp{1.2cm}p{1.2cm}}
        \hline\hline
         \multirow{2}{*}{Data} & \multicolumn{6}{c}{Best-fit values} & \multirow{2}{*}{$\chi_{\text{min}}^{2}$} & \multirow{2}{*}{BIC} \\
         \cline{2-7}
          & $h$ & $\Omega_{b,0}$ & $\Omega_{DM,0}$ & $\nu$ ($\times 10^{-2}$) & $\xi_{0}^{\nu}$ & $M$ & & \\
         \hline
         \multicolumn{9}{c}{$\Lambda$CDM model} \\
         SNe+BAO & $0.700\pm 0.005$ & $0.047\pm 0.001$ & $0.264\pm 0.009$ & $\cdots$ & $\cdots$ & $-19.36\pm 0.02$ & $1553.4$ & $1583.2$ \\
         SNe+BAO+CC & $0.699\pm 0.005$ & $0.047\pm 0.001$ & $0.262\pm 0.008$ & $\cdots$ & $\cdots$ & $-19.36\pm 0.02$ & $1569.4$ & $1599.3$ \\
         Joint & $0.702\pm 0.005$ & $0.047\pm 0.001$ & $0.262\pm 0.008$ & $\cdots$ & $\cdots$ & $-19.35\pm 0.01$ & $1707.3$ & $1737.1$ \\
         \hline
         \multicolumn{9}{c}{Dissipative UG model} \\
         SNe+BAO & $0.734\pm 0.010$ & $0.041\pm 0.001$ & $0.333\pm 0.019$ & $0.36\pm 9.75$ & $0.037\pm 0.017$ & $-19.24\pm 0.03$ & $1533.9$ & $1578.6$ \\
         SNe+BAO+CC & $0.720\pm 0.008$ & $0.043\pm 0.001$ & $0.306\pm 0.015$ & $0.84\pm 9.32$ & $0.025\pm 0.015$ & $-19.29\pm 0.02$ & $1557.5$ & $1602.3$ \\
         Joint & $0.723\pm 0.007$ & $0.042\pm 0.001$ & $0.309\pm 0.015$ & $-1.08\pm 9.58$ & $0.029\pm 0.015$ & $-19.28\pm 0.02$ & $1691.7$ & $1736.5$ \\
         \hline
         \multicolumn{9}{c}{Dissipative UG model with $\xi_{0}=1$} \\
         SNe+BAO & $0.736\pm 0.009$ & $0.041\pm 0.001$ & $0.335\pm 0.019$ & $0.06\pm 0.25$ & $0.040\pm 0.009$ & $-19.24\pm 0.03$ & $1564.6$ & $1601.8$ \\
         SNe+BAO+CC & $0.721\pm 0.007$ & $0.042\pm 0.001$ & $0.308\pm 0.016$ & $-0.02\pm 0.11$ & $0.028\pm 0.007$ & $-19.29\pm 0.02$ & $1582.5$ & $1619.8$ \\
         Joint & $0.724\pm 0.007$ & $0.042\pm 0.001$ & $0.312\pm 0.015$ & $-0.03\pm 0.14$ & $0.030\pm 0.007$ & $-19.28\pm 0.02$ & $1713.6$ & $1750.9$ \\
         \hline
         \multicolumn{9}{c}{UG model without dissipation} \\
         SNe+BAO & $0.719\pm 0.007$ & $0.044\pm 0.001$ & $0.315\pm 0.017$ & $19.60\pm 4.80$ & $\cdots$ & $-19.29\pm0.02$ & $1538.2$ & $1575.5$ \\
         SNe+BAO+CC & $0.710\pm 0.006$ & $0.044\pm 0.001$ & $0.297\pm 0.014$ & $14.84\pm 4.33$ & $\cdots$ & $-19.31\pm 0.02$ & $1559.5$ & $1596.9$ \\
         Joint & $0.714\pm 0.005$ & $0.044\pm 0.001$ & $0.301\pm 0.013$ & $16.13\pm 4.14$ & $\cdots$ & $-19.31\pm 0.02$ & $1695.1$ & $1732.4$ \\
         \hline\hline
    \end{tabularx}
    \caption{The best-fit values at the $1\sigma$ CL and the goodness-of-fit criteria for the $\Lambda$CDM, dissipative UG, and UG without dissipation models. These values were obtained through the MCMC analysis described in Section \ref{sec:Constraints}. Specifically, the dissipative UG model with $\xi_{0}=1$ requires $\xi_{0}^{\nu}=|\nu|^{1/2}$, which was estimated directly from the MCMC chains. The joint analysis corresponds to the combined data SNe+BAO+CC+GL+BHS.}
    \label{tab:best-fits}
\end{table*}

\begin{table}[ht]
    \centering
    \begin{tabularx}{\columnwidth}{p{3.4cm}YY}
        \hline\hline
        Model & $\Delta\chi_{\text{min}}^{2}$ & $\Delta$BIC \\
        \hline
        \multicolumn{3}{c}{SNe+BAO} \\
        Dissipative UG & $19.5$ & $4.6$ \\
        Dissipative UG ($\xi_{0}=1$) & $-11.2$ & $-18.6$ \\
        UG without dissipation & $15.2$ & $7.7$ \\
        \hline
        \multicolumn{3}{c}{SNe+BAO+CC} \\
        Dissipative UG & $11.9$ & $-3.0$ \\
        Dissipative UG ($\xi_{0}=1$) & $-13.1$ & $-20.5$ \\
        UG without dissipation & $9.9$ & $2.4$ \\
        \hline
        \multicolumn{3}{c}{Joint} \\
        Dissipative UG & $15.6$ & $0.6$ \\
        Dissipative UG ($\xi_{0}=1$) & $-6.3$ & $-13.8$ \\
        UG without dissipation & $12.2$ & $4.7$ \\
        \hline\hline
    \end{tabularx}
    \caption{The differences $\Delta\chi_{\text{min}}^{2}=\left(\chi_{\text{min}}^{\Lambda\text{CDM}}\right)^{2}-\left(\chi_{\text{min}}^{\text{model}}\right)^{2}$ and $\Delta\text{BIC}=\text{BIC}_{\Lambda\text{CDM}}-\text{BIC}_{\text{model}}$ between the $\Lambda$CDM and the UG models. By definition, a positive value indicates that the data favor the model under study, whereas a negative value implies evidence in favor of the $\Lambda$CDM model.}
    \label{tab:deltagoodness}
\end{table}

From Tables \ref{tab:best-fits} and \ref{tab:deltagoodness}, it can be observed that both the dissipative UG and UG without dissipation models exhibit lower $\chi^{2}_{\text{min}}$ values compared to the $\Lambda$CDM model for all dataset combinations considered. Only the dissipative UG model with $\xi_{0}=1$ is less competitive than $\Lambda$CDM in describing the aforementioned data. The BIC provides further insight; there is very strong evidence against the dissipative UG case with $\xi_{0}=1$ compared to $\Lambda$CDM across all datasets. On the other hand, for the dissipative UG gravity, we find evidence against $\Lambda$CDM when using SNe+BAO, but evidence against the UG model when using the SNe+BAO+CC joint analysis. However, when GL and BHS data are included, the BIC does not strongly favor either the dissipative UG or the $\Lambda$CDM model. Interestingly, the UG model without dissipation shows evidence against $\Lambda$CDM for the SNe+BAO+CC data and the joint analysis, with strong evidence against the $\Lambda$CDM model for the SNe+BAO data. This is a consequence of having one fewer free parameter compared to the dissipative UG case. Notably, this advantage does not extend to the dissipative UG with $\xi_{0}=1$ (despite having the same number of parameters as the non-dissipative model), indicating that the extra free parameter is necessary to distinguish the evolution of the bulk viscosity from the diffusion function, as seen in Eqs. \eqref{diffusion} and \eqref{bulkvicosity}. Consequently, both the dissipative UG and the UG without dissipation models remain competitive with the $\Lambda$CDM model in describing the combined SNe+BAO+CC+GL+BHS data. While the dissipative UG model consistently yields lower $\chi^{2}_{\text{min}}$	values, the UG model without dissipation is favored by the BIC in all cases due to its simplicity. Nevertheless, both models remain highly competitive when considering the SNe+BAO dataset alone.

Focusing our analysis on the best-fit values presented in Table \ref{tab:best-fits}, we observe that the dissipative UG model—in both general and particular cases—exhibits $H_{0}$ values closer to $H_{0}=73.04\pm 1.04\frac{km/s}{Mpc}$, as reported by Riess \textit{et al.} \cite{Riess:2021jrx} using Cepheids. Specifically, for the SNe+BAO dataset, the obtained $H_{0}$ is practically identical to the SH0ES value, which serves as a point in favor of the dissipative UG model. Conversely, in all UG variants, the best-fit value for $\Omega_{b,0}$ is slightly lower than in the $\Lambda$CDM case, while $\Omega_{DM,0}$	is consistently larger. This is a consequence of the additional components in the universe, which, by redistributing energy density, modify the overall expansion dynamics. Regarding the best-fit value for $\nu$, note that across all considered datasets, it remains close to zero but is non-negligible. Recalling from Eq. \eqref{diffusion} that $\nu=0$ recovers the $\Lambda$CDM model, the competitive goodness-of-fit criteria obtained for our model compared to $\Lambda$CDM are precisely due to this additional component. Interestingly, for the UG model without dissipation, the value of $\nu$ is larger than in the dissipative case, as it must account for the dynamics without the contribution of bulk viscosity. Furthermore, the poor goodness-of-fit criteria for the UG model with $\xi_{0}=1$ are a direct consequence of this fixed value. Since $\xi_{0}=1$ implies $\xi_{0}^{\nu}=|\nu|^{1/2}$, the value of $\nu$ is strictly linked to the viscosity to force $\xi_{0}^{\nu}$ toward values seen in the general dissipative case. While this could suggest that dissipation plays a more significant role than the diffusion function, we must highlight that in our model, dissipation is actually a consequence of the energy diffusion of the universe, according to our ansätze in Eqs. \eqref{diffusion} and \eqref{bulkvicosity}.

Finally, from Figs. \ref{fig:TriangleDissipativeUG} and \ref{fig:TriangleParticularUG}, we observe that for the dissipative UG cases, there is no clear preference for either positive or negative values of $\nu$. According to Eq. \eqref{diffusion}, this implies no definitive preference for the sign of the energy diffusion function. Notably, Fig. \ref{fig:TriangleParticularUG} reveals a multimodality in the distribution of $\nu$ around $\nu=0$ for the dissipative UG case with $\xi_{0}=1$. This is likely a consequence of the fixed-value constraint $\xi_{0}^{\nu}=|\nu|^{1/2}$, which couples the parameters in such a way that the likelihood surface develops multiple peaks. Although this multimodality could be broken by imposing a prior such as $\nu>0$ or $\nu<0$, the overall performance of this specific model compared to $\Lambda$CDM is so poor that it can be discarded. Conversely, Fig. \ref{fig:TriangleUG} shows that for the UG model without dissipation, the data favor $\nu>0$, implying a positive diffusion function.

%%%%%%%%%%%%%%%%%%%%%%%%%%%%%%%%%%%%%%%%%%%%%%%%%%%%%%%%%%%%
\section{\label{sec:DifussionAndBulk}Energy Diffusion as the Source of Cosmic Dissipation}
%%%%%%%%%%%%%%%%%%%%%%%%%%%%%%%%%%%%%%%%%%%%%%%%%%%%%%%%%%%%
Throughout this work, we have mathematically implemented the power-law ansatz $\xi=\xi_0|Q|^{1/2}$ to couple the bulk viscosity coefficient $\xi$ of the cosmic fluid with the energy diffusion function $Q$ inherent to the UG framework. While the above effective choice leads to an elegant analytical solution for the late-time Hubble parameter, $H(z)$, and yields excellent observational consistency, it is crucial to establish the physical and thermodynamic mechanisms that justify such a macroscopic relationship. In what follows, we provide some physical arguments supporting the plausibility that cosmological dissipation may naturally arise from energy non-conservation via diffusion.

Equation~(\ref{UGconservation}) shows that diffusion acts as an energy flux which, for $\dot{Q}>0$, feeds the energy density $\rho$ of the matter sector. This interpretation follows naturally once the matter sector is treated covariantly as an open system, in accordance with the modified conservation law obtained from the unimodular Noether identity given by Eq.~\eqref{noether}. This expression makes explicit that only the total current $T_{\mu\nu}-g_{\mu\nu}Q$ is divergence-free, while the matter component $T_{\mu \nu}$  alone is not. The term $g_{\mu\nu}Q$ thus plays the role of a ``geometric reservoir'' with which the matter sector exchanges energy. Schematically
\begin{equation}
\underbrace{T_{\mu\nu}}_{\text{matter (open)}}
\;+\;\underbrace{(-g_{\mu\nu}Q)}_{\text{geometric reservoir}}
\;=\;\underbrace{T_{\mu\nu}-g_{\mu \nu}Q}_{\text{closed total}}.
\end{equation}
Recognizing the matter sector as an open system in the gravitational sense allows one to define a covariant boundary separating it from the geometric reservoir. The energy exchange across this boundary is encoded in the well-defined, covariant flux $\dot{Q}$ threading the comoving worldlines, while the composite system $\{\text{matter}+\text{geometric reservoir}\}$ remains closed. This construction realises, in a precise covariant and geometrical form, a foundational principle of 
thermodynamics: open systems exchange energy with a reservoir; isolated systems do not.

At the level of cosmological phenomenology, the modified conservation equation~\eqref{UGconservation} is operationally indistinguishable from the following scenarios:
\begin{itemize}
\item[\textbf{i)}]\textbf{Particle production in curved spacetime}, in which 
$n^{\alpha}{}_{;\alpha}=\Gamma\,n$ effectively sources $\rho$ through the gravitational field (Parker--Hawking--Zel'dovich mechanism).
\item[\textbf{ii)}]\textbf{Interacting dark-sector models}, with an explicit coupling $Q_{\rm int}$, i.e.,
\begin{eqnarray}\label{Interacting}
\dot{\rho}_{m}+3H(1+w_{m})\rho_{m}&=-Q_{\rm int},\nonumber\\
\dot{\rho}_{x}+3H(1+w_{x})\rho_{x}&=+Q_{\rm int}.
\end{eqnarray}
\item[\textbf{iii)}]\textbf{Running-vacuum cosmologies}, scenarios where $\rho_{\rm vac}=\Lambda+\nu H^{2}$, in which the cosmological term slowly drains into matter (or vice versa) along the cosmic evolution.
\end{itemize}

Once the openness of matter from the gravitational standpoint is accepted, the relevant question is no longer ``whether viscosity should be added by hand'', but rather ``what is the minimal hydrodynamic response of the matter fluid to the flux $\dot{Q}$ crossing its boundary''. In an isotropic and homogeneous universe, the only scalar response available to a relativistic fluid near equilibrium is a bulk-viscous pressure $\Pi=-3H\xi$. As a consequence, the second law of thermodynamics requires that the total entropy production satisfy the generalized condition
\begin{equation}\label{EntropyProduction}
T\,S^{\alpha}{}_{;\alpha}=-3H\Pi-\dot{Q}=9H^{2}\xi-\dot{Q}\;\geq\;0,
\end{equation}
which represents a consistency constraint on the model parameters. The fluid, therefore, cannot re-equilibrate instantaneously: the diffusion flux $\dot{Q}$ drives it away from equilibrium and is balanced, at leading order, by a bulk-viscous response that generates entropy. In this sense, dissipation does not have to be postulated independently; it emerges as the minimal thermodynamically consistent reaction of the matter fluid to the geometric energy flux.

%%%%%%%%%%%%%%%%%%%%%%%%%%%%%%%%%%%%%%%%%%%%
\section{\label{sec:Conclusions}Conclusions}
%%%%%%%%%%%%%%%%%%%%%%%%%%%%%%%%%%%%%%%%%%%%
In this paper, we have discussed the late-time behavior of a novel FLRW cosmological model within the UG framework, where the dominant fluid is a pressureless matter, composed of baryons and CDM, which experiences energy diffusion and bulk viscosity. For the energy diffusion function, $Q$, inherent to UG, we proposed the ansatz $Q=\nu H^{2}$, while for the bulk viscosity $\xi$, we used the expression $\xi=\xi_{0}|Q|^{1/2}=\xi_{0}^{\nu}|H|$, where $\nu$ and $\xi$ are dimensionless parameters. As previously discussed, our choice for $Q$ and $\xi$ means that the dissipative process in the main fluid can be interpreted as a consequence of the additional energy due to the diffusion function of the UG framework. Even more, given these ansätze, and considering that the vacuum energy density is time-dependent by construction in the UG models, we obtain a vacuum evolution $\rho_{\text{vac}}\equiv\Lambda+Q=\Lambda+\nu H^{2}$. This result is formally similar to the one obtained in the framework of perturbative Quantum Field Theory in a curved classical background, with a general expression $\rho_{\text{vac}}(H,\dot{H})=\Lambda+\nu H^{2}+\tilde{\nu}\dot{H}$~\cite{Sola:2015rra}, which has been recently tested using BAO data from DESI DR1 in the context of $f(Q)$ gravity, demonstrating that it remains competitive with the $\Lambda$CDM model~\cite{universe12010025}. Our model corresponds to the case where $\tilde{\nu}=0$, although it is important to emphasize that both approaches are not strictly equivalent.

Under the above ansätze, we obtain an analytical solution that is confronted with the observational data by performing a robust MCMC analysis using a comprehensive suite of late-time observations, including SNe Ia, CC, BAO, GL, and BHS data. Our results demonstrate that both the general dissipative UG model and the UG model without dissipation are highly competitive against the standard $\Lambda$CDM model. Specifically, these models consistently yield lower $\chi^{2}_{\text{min}}$ values across all dataset combinations considered, indicating a superior ability to fit the observational data. While the dissipative model provides the best overall fit, the BIC identifies the UG model without dissipation as the most favored candidate due to reduced parameter space.

One of the most significant findings of our analysis is that the diffusion parameter $\nu$ is non-negligible across all datasets, indicating that the observations favor a departure from the $\Lambda$CDM dynamics (where $\nu=0$). This result suggests that a very small but nontrivial energy nonconservation is compatible with—and potentially preferred by—late-time cosmological data. In the non-dissipative case, a positive diffusion function is clearly preferred by the data. We also found that the particular UG case with $\xi_{0}=1$ leads to a poor goodness-of-fit and multimodality in the posterior distributions, suggesting that the energy diffusion and bulk viscosity should be treated as independent degrees of freedom to correctly capture the expansion history of the universe. The observed shifts in the matter density parameters, $\Omega_{b,0}$ and $\Omega_{DM,0}$, further highlight the active role of the additional UG components in modifying the overall expansion dynamics.

To clarify why the dissipative and non-dissipative UG models are favored by the data, it is helpful to consider, as a first approximation, that this preference arises from the nature of UG itself. From Eq. \eqref{lambda}, the term $\lambda=\Lambda+Q$ can be interpreted as a variable CC, a concept that has been extensively explored in the literature as a motivated alternative to the standard $\Lambda$ term~\cite{Dymnikova:2001jy,Mukhopadhyay:2007ed}. While we have already discussed this in the context of the running vacuum to justify our choice for the diffusion function, it is important to understand that this interpretation holds for any choice of $Q$, not just the one adopted in this study. On the other hand, while one might argue that the lower $\chi_{\text{min}}^{2}$ of the dissipative UG model is merely due to the additional free parameters, it may also be a direct consequence of the dissipation itself. By rewriting Eqs. \eqref{UGfluids} and \eqref{UGpressures} in a form similar to their GR counterparts (Eqs. \eqref{GRfluids} and \eqref{GRpressures}), we can define the effective CC with an effective barotropic index
\begin{equation}\label{CCeffective}
    \omega_{\text{eff}}=\frac{p_{\text{eff}}}{\rho_{\text{eff}}}\equiv-\frac{\Lambda+Q-\Pi}{\Lambda+Q}=-1+\frac{\Pi}{\Lambda+Q},
\end{equation}
where the barotropic index evolves over time. This evolution depends on the specific choices for the diffusion function $Q$ and the bulk viscosity $\xi$, given that $\Pi=-3H\xi$ in Eckart’s theory. This behavior is particularly interesting as the DESI DR2 Results II show a tension with CMB-derived values, which is alleviated by the $\Lambda$CDM model with a DM component with a time-evolving EoS~\cite{DESI:2025zgx}. While a deeper exploration of these considerations is warranted, it remains the subject of future work, as our current aim is to present the novel framework of dissipative UG and provide a preliminary study of its capabilities.

In conclusion, UG with dissipation offers a compelling and statistically sound alternative to the standard model, demonstrating strong consistency with diverse observational probes, most notably the recent DESI DR2 BAO sample. Although the dissipative model remains as competitive as $\Lambda$CDM from a BIC perspective, it offers the distinct advantage of a significantly lower $\chi_{\text{min}}^{2}$ value compared to the non-dissipative case. Furthermore, it yields a best-fit for $H_{0}$ in excellent agreement with the model independent value $H_{0}=73.04\pm 1.04\frac{km/s}{Mpc}$ reported by Riess \textit{et al.} \cite{Riess:2021jrx}. Therefore, considering that this framework is built upon a theory that inherently alleviates the CC problem, it emerges as a robust and theoretically well-motivated alternative to the $\Lambda$CDM paradigm.

%%%%%%%%%%%%%%%%%%%%%%%%%%%%%%%%%%%%%%%%%%%%%%%%%%%%%
\section*{\label{sec:Acknowledgments}Acknowledgments}
%%%%%%%%%%%%%%%%%%%%%%%%%%%%%%%%%%%%%%%%%%%%%%%%%%%%%
This work is dedicated to the memory of Dr. Jose Jovel, who passed away during the development of this research. N.C. and G.P. acknowledge financial support from the Chilean National Agency for Research and Development (ANID) through Fondecyt Grant No. 1250969, Chile. E.G. acknowledge Vicerrectoría de Investigación y Desarrollo Tecnológico (VRIDT) at Universidad Católica del Norte (UCN) for the scientific support provided by Núcleo de Investigación en Simetrías y la Estructura del Universo (NISEU-UCN), Resolución VRIDT N°200/2025.

\bibliographystyle{apsrev4-1}
\bibliography{bibliography}
\end{document}